\documentclass[journal=jpcbfk,manuscript=article,layout=twocolumn]{achemso}
\usepackage{graphicx}
\usepackage{amsmath}
\usepackage{multirow}
\usepackage{lipsum}
\usepackage{float}
\usepackage{xcolor}

\newcommand{\sub}[1]{\ensuremath{_{\textrm{#1}}}}
\newcommand{\super}[1]{\ensuremath{^{\textrm{#1}}}}
\renewcommand{\vec}[1]{\ensuremath{\textbf{#1}}}

\newcommand{\Lagr}{\mathcal{L}}
\setcitestyle{super}

\title{Machine Learning Models Capture Plasmon Dynamics in Ag Nanoparticles}

\def\LANL{Theoretical Division, Los Alamos National Laboratory, Los Alamos, NM 87545, USA}
\def\LANLCCS{Computer, Computational and Statistical Sciences (CCS) Division, Los Alamos National Laboratory, Los Alamos, NM 87545, USA}
\def\LANLCINT{Center for Integrated Nanotechnologies Los Alamos National Laboratory, Los Alamos, NM 87545, USA}

\setcitestyle{super}
\author{Adela Habib}\email{ahabib@lanl.gov}\affiliation{\LANL}
\author{Nicholas Lubbers} \affiliation{\LANLCCS}
\author{Sergei Tretiak}\affiliation{\LANL}\alsoaffiliation{\LANLCINT}
\author{Benjamin Nebgen}\email{bnebgen@lanl.gov}\affiliation{\LANL}
\date{\today}

\let\oldmaketitle\maketitle
\let\maketitle\relax

\begin{document}

\twocolumn[
\begin{@twocolumnfalse}
\oldmaketitle
\vspace{-0.2in}
\begin{abstract}
Highly energetic electron-hole pairs (hot carriers) formed from plasmon decay in metallic nanostructures promise sustainable pathways for energy-harvesting devices. However, efficient collection before thermalization remains an obstacle for realization of their full energy generating potential. Addressing this challenge requires detailed understanding of physical processes from plasmon excitation in metal to their collection in a molecule or a semiconductor, where atomistic theoretical investigation may be particularly beneficial. Unfortunately, first-principles theoretical modeling of these processes is extremely costly, preventing a detailed analysis over a large number of potential nanostructures and limiting the analysis to systems with a few 100s of atoms. Recent advances in machine learned interatomic potentials suggest that dynamics can be accelerated with surrogate models which replace the full solution of the Schr\"{o}dinger Equation. Here, we modify an existing neural network, Hierarchically Interacting Particle Neural Network (HIP-NN), to predict plasmon dynamics in Ag nanoparticles. The model takes as minimum as three time-steps of the reference real-time time-dependent density functional theory (rt-TDDFT) calculated charges as history and predicts trajectories for 5 femtoseconds in great agreement with the reference simulation. Further, we show that a multi-step training approach in which the loss function includes errors from future time-step predictions, can stabilize the model predictions for the entire simulated trajectory ($\sim$25~fs). This extends the model's capability to accurately predict plasmon dynamics in large nanoparticles of up to 561 atoms, not present in training dataset. More importantly, with machine learning models we gain a speed-up of $\sim$200 times as compared with the rt-TDDFT calculations when predicting important physical quantities such as dynamic dipole moments in Ag$\sub{55}$ and $\sim$4000 times for extended nanoparticles that are 10 times larger. This underscores the promise of future machine learning accelerated electron/nuclear dynamics simulations for understanding fundamental properties of plasmon-driven hot carrier devices. 
\end{abstract}
\end{@twocolumnfalse}
\vspace{0.2in}
]

\maketitle
\section{Introduction}
Plasmons are collective oscillations of free electrons in metal nanostructures. Their decay enables formation of highly energetic electrons and holes known as hot carriers. These plasmonic hot carriers promise useful applications in photodetection, \cite{Tagliabue2020CuPhotoDetection} photovoltaics, \cite{Pillai2010PlasmonicPhotovoltaic} and photocatalysis. \cite{Cortes2022PlasmonPhotoChemRev, zhang2021PlasmonPECRev} However, experimental measurements have shown low photon to carrier conversion efficiencies in these devices.\cite{Dorodnyy2018ConversEfficiencyRev, Zhang2022HotCarrEfficiencyReview}
Understanding the root cause of this low rate of carrier collection requires scrutinizing various quantum-mechanical (QM) processes including the excitation of plasmons, plasmonic decay into generating hot carriers, the transport of hot carriers in the metal, and finally their collection across the metal-semiconductor or metal-molecule interface. Many factors including geometry, electronic band structure of the system, energy and momentum matching at the interface contribute to these processes. However, studying these processes only by experimental means is limited due to a variety of challenges such as interpretation of experimental data and time/spatial resolution of specific experiments. \cite{Reddy2021HotCarrierDevEfficiency, Kumar2021HotCarrierReview}

Theoretical studies of hot carriers have also proven extremely challenging due to the  prohibitive computational cost of simulating these processes from first principles.\cite{Kumar2021HotCarrierReview} To alleviate this cost, several studies have employed theoretical models that take classical and/or semi-classical approach. For example, in the so called Jellium model, it is customary to assume a free-electron-like gas with a positive charge background in metals, and use perturbation theory such as Fermi's golden rule to estimate hot carrier generation rates. \cite{Zhang2014HotCarrierGovorov, Manjavacas2014HotCarrierJellium}
A more advanced but expensive approach is using density functional theory (DFT) to evaluate electronic and phonon properties including electron-electron (e-e) and electron-phonon (e-ph) interaction matrix elements, and employ Fermi's golden rule along with Boltzmann transport equations (BTE) to estimate hot carrier generation and their transport in metals. \cite{PhononAssisted, Sundararaman2014DirectTransitions, Brown2017HotCarrBoltzmanTransport} Calculations from these methods have provided valuable insights into hot carrier dynamics that validate some aspects of experimental measurements, for example, estimating the percentage of hot hole transfer from Au to p-GaN semiconductor. \cite{Tagliabue2020HoleTransferDFT} However, atomistic DFT approach is only computationally efficient for unit cell calculations and is costly for realistic device systems that include interfaces. Moreover, it neglects the underlying geometric details of the nanoparticles that influence electronic band structure, affecting plasmon excitation, hot carrier generation and their transfer across a desired interface. \cite{Tagliabue2018HotCarrierIQE}

Bound by the restrictions related to computational resources, some studies have nevertheless accounted for the underlying geometric factors in plasmon dynamics simulation by using methods such as real-time density functional theory (rt-TDDFT) and non-adiabatic molecular dynamics (NAMD).\cite{Ma2015AtomisticRttddftSilver55, Run2014GoldTitaniaNAMD, Douglas2019HCrtTDDFTAuAgNanoparticles, Douglas2016PlasmonAgCappedWater} These studies evolve electronic wave functions over time for tens of femtoseconds, and have provided useful insights, separating the plasmonically generated hot carriers from the resonant single-particle excitations created by direct photon absorption. However, application of these methods is still hindered by their enormous computational cost even for nanoparticles of sizes approximately 3~nm, thus making them less suitable for a high-throughput screening of systems with desirable material and geometric properties for efficient plasmon-driven hot carrier devices.  

Unlike the QM methods' poor computational scaling with system size, machine learning (ML) methods typically scale linearly and have successfully accelerated dynamics calculations in a broad variety of atomistic sciences. For example, molecular dynamics (MD) simulations of atomic trajectories with ML interatomic potentials is now possible for large systems ($\sim$millions of atoms) with reasonable computational expense. \cite{Schutt2017NNforPE, Schutt2018SchNet, Behler2017NNPEnergy, Yao2018TensorMolNNPotEn, Smith2019ANI-1x, Lubbers2018HIPNN, Kulichenko2021NNReview, Smith2021NNPotentialAluminum} Moreover, atomistic neural networks (NNs) can capture non-local charge transfer between atoms in large systems, resulting in a DFT-level accurate potential energies.\cite{Ko20214thOrderBPChargeMLtrain, Sifain2018transferChargeML, Nebgen2018ChargesHIPNN, Zubatyuk2021MoleculesElectron, Zubatyuk2019AIMNETEnergiesCharges} 
Notably, currently published models assume that the electronic configuration is either purely a function of local atomic environment or allowed to equilibrate through charge equilibration schemes. Consequently, none of these schemes is useful for simulating hot carrier dynamics, which are manifestly out-of-equilibrium.

In this paper we extend the capability of Hierarchically Interacting Particle Neural Network (HIP-NN),\cite{Lubbers2018HIPNN} an existing atomistic NN model, to track charges during plasmon resonance in nanoparticles. The ability to track non-equilibrium electron movement at ML speeds facilitates new insights into a variety of applications including light-matter interactions, non-equilibrium processes, chemical reactions, etc. The accurate modeling of these effects at large length and time scales are critical to the design of efficient devices in many technological fields. In the field of plasmonic hot-carriers particularly, this means ML models accelerating predictions of carrier dynamics including plasmon excitation, their decay into hot carriers, and finally the charge-transfer process across a chosen interface of interest. In this chain of processes, the first two, i.e., plasmon excitation and decay into hot carrier distributions play definitive roles in determining the initial device performance.  

This paper applies our data-based approach for atomistic predictions of electron dynamics to silver nanoparticles of icosahedral and octahedron shapes using HIP-NN. With a history-based characteristic, the ML model is trained on induced Mulliken charge trajectories out to 1300 time steps in nanoparticles of Ag$\super{Icos}\sub{55,147}$ and Ag$\super{Octa}\sub{146}$. These time-dependent trajectories are initiated by a short optical pulse triggering long-time plasmonic coherent charge oscillations in the silver nanoparticles. The trained models show remarkable performance, predicting derived quantities such as dynamic dipole moments that show excellent quantitative agreement with the reference rt-TDDFT calculations for $\sim$5~fs after an initial excitation pulse. To further refine the HIP-NN models, a multi-step training model is developed where losses are back propagated through successive predictions made by the model. This further improves the model performance not only for the nanoparticles in the training set but also for extended nanoparticles of Ag$\super{Icos}\sub{309}$, and Ag$\super{Icos}\sub{561}$. Additionally, in a similar approach, we also train a model on positive and negative charges as holes and electrons on an atom to formulate an ML model for propagating hot carriers. The resulting model qualitatively captures these carriers behavior for the entire trajectory (1300 time steps), but suggests that a more refined approach using orbital-based descriptors of hot carriers may improve the performance. Overall, we find that a history-based ML model can
predict charge trajectories and derived physical quantities such as dynamic dipole moments with great accuracy in nanoparticles of 55 to 561 atoms, giving 200 to 4000 times speed-up against the rt-TDDFT simulation results. This underscores the computational efficiency and applicability of our model to realistic device sizes, providing a pragmatic tool to rapidly gain fundamental insights about hot carrier dynamics. 

\section{Methods}
\subsection{Data generation}
\label{subsec:DataGen}
To model electron dynamics in nanoparticles with machine learning, a reference training dataset is required. This dataset is constructed from time-series data describing charge trajectories. For this, we use QM simulation tools that can capture the dynamics of charges in metals after their irradiation by a time-dependent laser pulse. We choose linear response rt-TDDFT, which has been used for studying plasmon dynamics in nanoparticles. \cite{Kuisma2015rtTDDFTLCAO, Ma2015AtomisticRttddftSilver55} Here, we briefly describe the dynamics as implemented in the open source software GPAW, \cite{Mortensen2005GPAWOriginal, Kuisma2015rtTDDFTLCAO} and the calculation of the training data. 

GPAW solves time-dependent Kohn-Sham equation in the partial augmented wave (PAW) formalism,
\begin{equation}
    \frac{\partial \psi_n (\vec{r}, t)}{\partial t} = H_\mathrm{KS} (t) \psi_n (\vec{r},t),
\label{eq:KStddft}
\end{equation}
where the time-dependent Kohn-Sham Hamiltonian $H_\mathrm{KS} = H^{(0)}_\mathrm{KS} + \vec{E}_{0} \text{sin}(\omega_{p} (t-t_c)) \mathrm{e}^{(-(t-t_c)^2/(2\tau^2_0))}$, consists of the ground state single particle term and the Guassian pulse perturbation. Here, $\omega_p$ is the plasmon frequency obtained from literature for each nanoparticle,\cite{Rossi2017KSrtTDDFTLCAOAg} $t_c$ is the pulse center in time set to 2.5~fs, $\tau_0$ is the pulse width set to 5.5~fs. The $\vec{E}_{0}$ represents the external electric field direction and magnitude. The eigen wave functions are represented as a linear combination of atomic orbitals (AO), the so-called (LCAO) basis, \cite{Kuisma2015rtTDDFTLCAO} 

\begin{align}
     \psi_n (\vec{r}, t) = \sum_\mu \phi_\mu (\vec{r}) C_{\mu n} (t),
\end{align}
where $\mu$ is the atomic orbital number and $n$ represents the Kohn-Sham energy state. This approach essentially reduces Eq.~\eqref{eq:KStddft} to propagating the coefficients $C_{\mu n} (t)$ in time. GPAW uses the semi-implicit Crank-Nicolson numerical method to solve the equation. For details, please see Ref.~\citenum{Kuisma2015rtTDDFTLCAO}.

\begin{figure}
\includegraphics[width=\columnwidth]{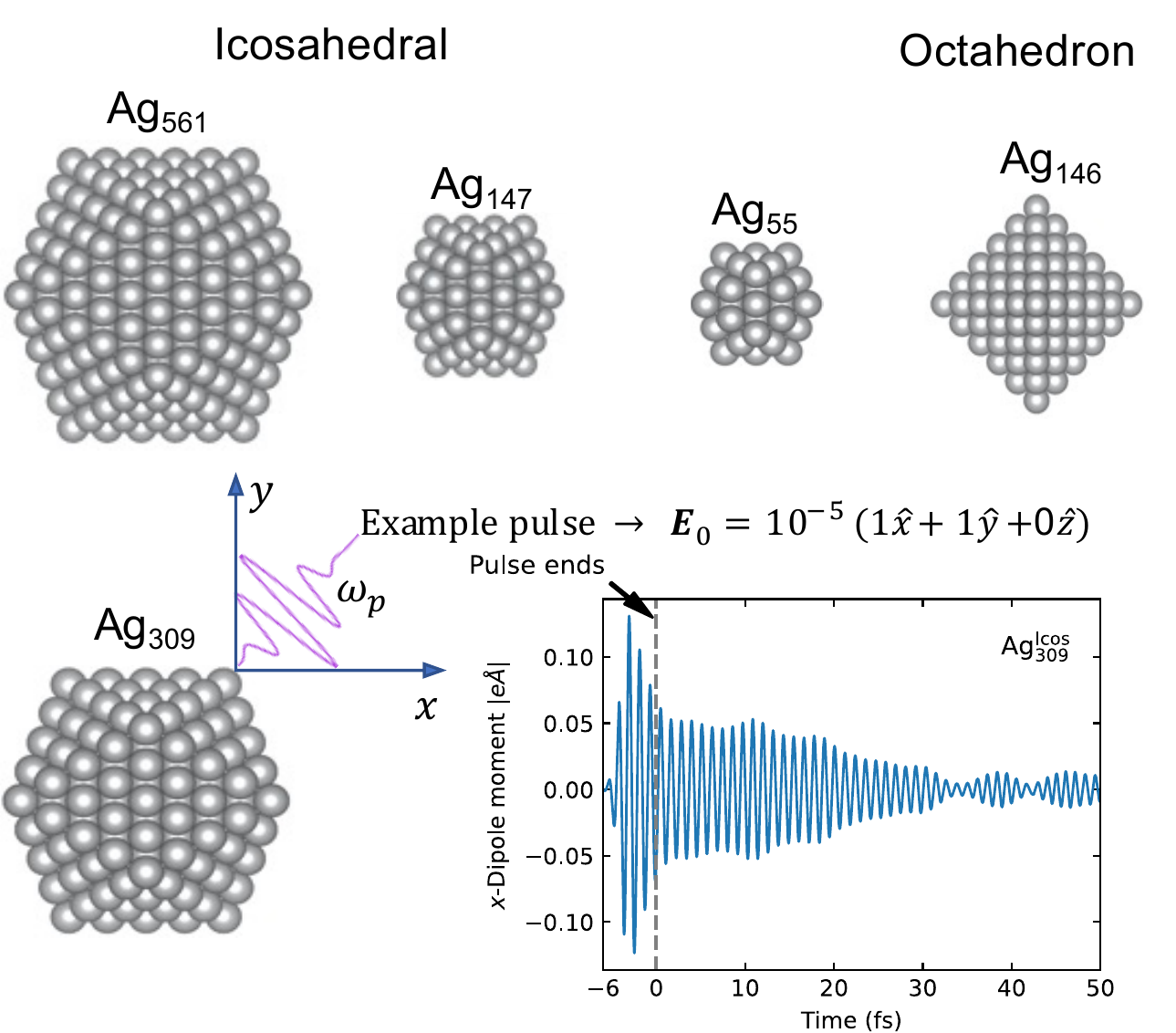}
\caption{Training data generation. The dataset includes nanoparticles of various sizes (55 - 561 atoms) and geometrical icosahedral and octahedron shapes. The trajectories of charges (electrons/holes) in these clusters show variation in incident pulse direction, magnitude, plasmon resonance frequencey and the nanoparticles positions. Induced dynamic dipoles result from plasmon oscillations over time that will vanish as plasmons decay to hot carriers as shown in a sample randomly chosen here.}
\label{fig:trainData}
\end{figure}

Figure~\ref{fig:trainData} shows a variety of silver nanoparticles contained in the dataset and a sample trajectory of Ag$\super{Icos}\sub{147}$. Charge-neutral nanoparticles of atom sizes in range 55 to 561, and shapes icosahedral and octahedron shapes are included. DFT calculations on the nanoparticles are performed with the Gritsenko-van Leeuwen-van Lenthe-Baerends solid-correlation (GLLB-sc) exchange correlation potential\cite{Kuisma2015rtTDDFTLCAO} and the LCAO basis set from 
Ref.~\citenum{Rossi2015LCAOBasisSet}, where this method was shown to accurately simulate $d$-electron states. The geometry of each nanoparticle is initially optimized using ground state DFT. With the perturbative Hamiltonian of time-dependent Gaussian pulse, the wave functions are then propagated with time steps of $dt$=20 attoseconds for 32 fs. To diversify the sampling of our dataset, we further randomize the incident electric field direction and the field strength to random numbers in the $10^{-5} - 10^{-3}$~V/$\text{\AA}$ range. It is important that all training of ML models is performed using reference rt-TDDFT trajectories after the optical pulse.

GPAW yields an electronic wave functions $\psi_n (\vec{r}, t)$ over time. Additionally, GPAW can calculate dynamic dipole moments as shown in Fig.~\ref{fig:trainData}. To convert the electronic wave function into a quantity suitable for machine learning applications, we compute the Mulliken charge for each atom. \cite{Mulliken}  While it is known that the Mullikan charge partitioning scheme is not the most physical, for this particular application, the precise partitioning scheme is unimportant. This is because physical quantities of interest in this study like dynamic dipole moment, are is an aggregate quantity, which are not highly dependant on precise charge partitioning between neighboring atoms.
We obtain these Mulliken charges by post-processing the wave functions to first get time-dependent density matrix operator in the LCAO basis,
\begin{align}
     \rho_{\mu\nu} (t) = \sum_n C_{\mu n} (t) f_n C^{\star}_{\nu n} (t),
\label{eq:densityMatKS}
\end{align}
where $f_n$ is the Fermi occupation factor, being 2 and 0 for occupied and virtual empty states, respectively. We can obtain the charge per atom $i$ (Mulliken charge population) by using the standard expression, 
\begin{align}
     q_i (t) = \sum_{\mu \in \textrm{atom} i} (\rho(t) S)_{\mu\mu},
\label{eq:Mulliken}
\end{align}
where $S_{\mu\nu} = \int d\vec{r} \phi_\mu (\vec{r}) \phi^{\star}_\nu (\vec{r})$ is the basis overlap matrix. We focus on induced charges
\begin{equation}
\delta q_{i}(t) = q_{i}(t) - q_{i}(t\sub{eq}),
\end{equation}
where $t\sub{eq}$ corresponds to the time before the nanoparticle is irradiated by a laser pulse to excite plasmons. Induced charges are typically far smaller, $\delta q_i \sim 10^{-3} e$, than the total Mulliken charge, as a result of the amplitude of incident laser pulse. 

Utilizing the time dependent charges, it is possible to construct the time dependent dipole moment of the nanoparticle. This provides a method to examine macroscopic charge behaviors that occur across the nanoparticle and investigate whether plasmon resonance is captured by the ML model. The instant calculation of dynamical dipole moment is given by the expression 
\begin{align}
    \Delta \boldsymbol{\mu}(t) = \sum^{N\sub{atoms}}_i \delta q_i(t) \vec{R}_i.
\end{align}

We also investigate the dynamics of hot carriers by tracking electron and hole differences separately. Most theoretical studies use an energy-space analysis for plasmonic hot carriers, for example, hot carrier distributions and scattering lifetimes as functions of their energy,\cite{PhononAssisted} or evolving these hot carrier energies and densities over time. \cite{Run2014GoldTitaniaNAMD, Douglas2019HCrtTDDFTAuAgNanoparticles, Zheng2019realTimeTDDFTAuGANhotCarrierInjection, Lu2022AuWSe2NAMDeph}. In this work, we use a simplified version of this analysis, where changes in occupation above (below) the Fermi level are assigned as electrons (holes), resulting in gross atomic assignment of negative (positive) charges on each atom.
We get the transition probability of electron from an occupied Kohn-Sham state $o$ to an unoccupied Kohn-Sham state $u$, creating a hole, from the change in density matrix, \cite{Rossi2020rtTDDFTHotCarrierGeneration} 
\begin{align}
    P_{ou} (t) = \left|\frac{\delta \rho_{ou} (t)}{\sqrt{f_o - f_u}} \right|^2,
\end{align}
where $f_{o (u)}$ is the occupation number associated with the electronic
states of $o$ ($u$) below (above) the Fermi energy level. Here, $\rho_{ou} (t)$ is given in Kohn-Sham basis transformed from the density matrix in AO basis per Eq.~\ref{eq:densityMatKS}. Please see Ref.~\citenum{Rossi2017KSrtTDDFTLCAOAg} for more details about this transformation. The change in density matrix is evaluated as $\delta \rho = \rho (t) - \rho (t\sub{eq})$ where $\rho (t\sub{eq})$ denotes the equilibrium density matrix before laser perturbation. The probabilities of holes and electrons for states $o$ (below) and $u$ (above) the Fermi level, respectively, are calculated as \cite{Rossi2020rtTDDFTHotCarrierGeneration}
\begin{align}
    P\super{h}_{o} (t)= \sum^{f_o > f_u}_{u} P_{ou} (t) \\
    P\super{e}_{u} (t)= \sum^{f_o > f_u}_{o} P_{ou} (t).
\end{align}
To convert these probabilities to an atom-level quantity, we use the Mulliken charge partitioning scheme per Eqs.~(\ref{eq:densityMatKS}-\ref{eq:Mulliken}), and calculate electron and hole populations as, 
\begin{align}
    \rho\super{h(e)}_{\mu\nu} (t) = \sum_{o (u)} C_{\mu o (u)} (t) P\super{h(e)}_{o (u)} (t) C^{\star}_{\nu o (u)} (t), \\
    \delta q\super{h(e)}_i (t) = \sum_{\mu \in \mathrm{atom}~i} (\rho\super{h(e)}(t) S)_{\mu\mu}.
\end{align}
Here, the equilibrium Fermi-Dirac distribution is essentially substituted with the electron-hole probabilities that are non-Fermi distribution-like. 

\subsection{Neural network structure}
\label{subsec:HIPNN}
The training targets for electron dynamics are induced Mulliken charges $\delta q_i(t)$ at the present time, given the history of all other atoms at a set of previous times. Denoting the collection of all atomic charges at time $t$ as $q(t)$ with no subscript, we denote the charge history of the system \emph{before} time $t$ as $\delta q(t:n\sub{hist})= \big(\delta q(t-n\sub{hist}),...,\delta q(t-1)\big)$, that is, the concatenated array of charges over an interval of $n_\mathrm{hist}$ time steps. Specifically, we want to use neural networks to predict charges on every atom over time as
\begin{align}
     \delta q(t)  =  \mathcal{NN} \big(\vec{R}, \delta q(t:n\sub{hist}) \big)
\end{align}
where $\vec{R}$ is the collection of nuclei positions. To implement the neural network model, we use the HIP-NN.\cite{Lubbers2018HIPNN} Here, we briefly review the architecture and explain the modifications made to facilitate the prediction of charge and electron/hole dynamics. 

HIP-NN can be viewed both as a message-passing neural network, which updates atomic features based on the information from nearby atoms, or as a continuous convolutional neural network, which updates atoms based on the local spatial distribution of features. The network converts atomic coordinates into pair-wise inter-atomic distances which HIP-NN uses as inputs, making it invariant to symmetry operations such as translation, rotation, and permutation. It also incorporates a locality constraint, such that individual operations in the network only take into account distances $r_{ij} < R_{\mathrm{cut}}$, for some user-defined radius $R_\mathrm{cut}$. Although in other works, the initial features are based on the atomic species, in this work, the input features are the time history of induced charges, that is, $\delta q(t:n_\mathrm{hist})$. 

\begin{figure*}
\includegraphics[width=\textwidth]{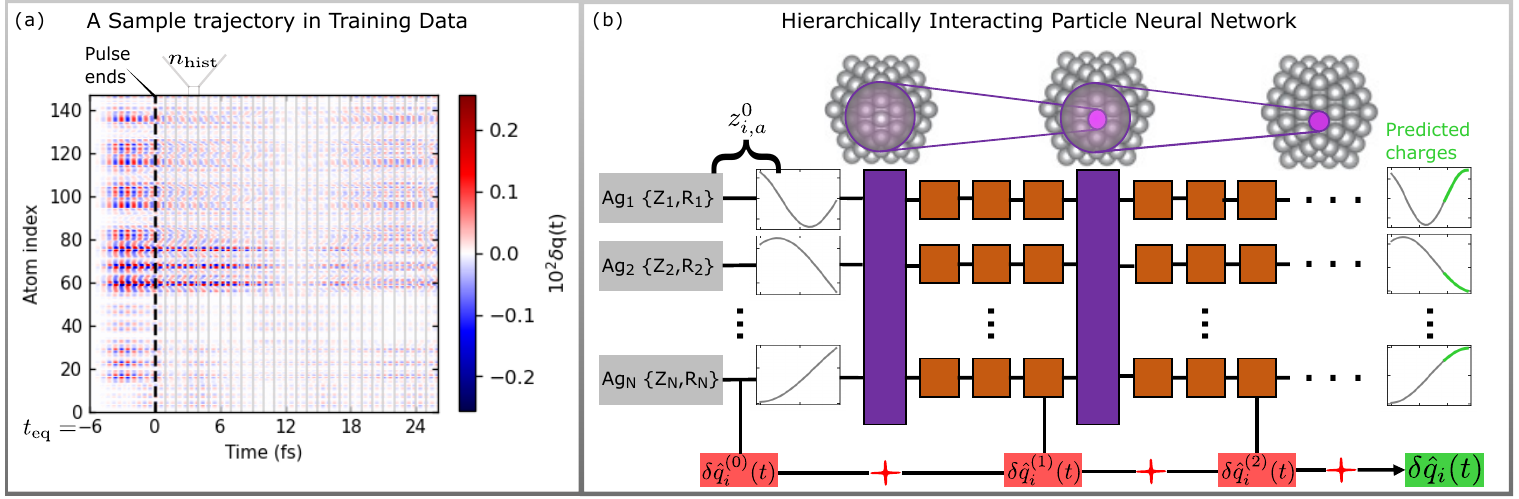}
\caption{A description of the training data and prediction task of HIP-NN. (a) shows a sample trajectory in a Ag$\super{Icos}\sub{147}$, where the first 6~fs ($t\sub{eq}$ = -6~fs to $t$ = 0) is the nanoparticle under the influence of an external electric field. At $t$ = 0~fs, the oscillating electric field is turned off and the electrons are allowed to relax for the remainder of the trajectory. (b) HIP-NN is given as inputs $n\sub{hist}$ charges starting at $t$ = 0 and asked to predict the next charge in the time sequence. In this way, HIP-NN can be used to propagate charges recursively for long time scales.}
\label{fig:HIPNNModel}
\end{figure*}

Figure~\ref{fig:HIPNNModel} shows how data is input into the network. Figure~\ref{fig:HIPNNModel}~(a) shows a sample trajectory of induced Mulliken charges over time for atoms in Ag$\super{Icos}\sub{147}$. For training we consider the trajectories immediately after the pulse ends, excluding the pulse region. Therefore, $t$ = 0 corresponds to the first time-step after the pulse ends, after which we divide the trajectory into intervals based on the variable parameter $n\sub{hist}$, shown as grey vertical lines. As shown in Fig.~\ref{fig:HIPNNModel}~(b), the time history for each atom is input to the network. HIP-NN processes information as arrays known as feature vectors, $z^{l}_{i,a}$ at each layer $l = 0, 1, ..., N\sub{layers}$; $z^{0}_{i,a}$ corresponds to the input values for each atom. The $a = 1, ..., N^{l}\sub{feature}$ indexes features, that is, individual neurons in  a layer, and $i = 1, ..., n\sub{atoms}$ indexes the atoms in the system. In subsequent layers these $z^{l}_{i,a}$ are updated using pair-wise distances  $r_{ij}$ and previous layers features $z^{l-1}_{i,a}$ to process information about the environment of each atom.

The HIP-NN neural network architecture consists of two types of layers: the on-site layers (dark orange boxes in Fig.~\ref{fig:HIPNNModel} (b)) and the interaction layers (dark purple boxes in Fig.~\ref{fig:HIPNNModel} (b)). The on-site layers are made of fully connected layers processing a single atom's features $z^{l}_{i,a}$,

\begin{align}
    z^{l+1}_{i,a} = f \left ( \sum_b W^{l}_{ab} z^{l}_{i,b} + b_{a}^{l} \right),
\end{align}
where $W^{l}_{ab}$ and $b^{l}_{a}$ are learned parameters and the activation function $f(x) = \mathrm{SoftPlus}(x) = \log(1+e^x)$. 

Unlike the on-site counterparts, the interaction layers (dark purple boxes in Fig.~\ref{fig:HIPNNModel}) not only operate on single atoms, but also consider interactions between an atom and its neighbors. This interaction is captured by including learnable parameters $v^{l}_{ab} (r_{ij})$, dependent on the pairwise distance. The update equation for interaction layers is as follows:
\begin{align}
    z^{l+1}_{i,a} = f \left ( \sum_{j,b} v^{l}_{ab} (r_{ij}) z^{l}_{j,b} +
    \sum_b W^{l}_{ab} z^{l}_{i,b} + b_{a}^{l} \right).
\end{align}
To represent $v_{ab}(r)$, it is expanded on a basis of sensitivity functions as
\begin{align}
    v^{l}_{ab} {(r)} = \sum_{\nu} V^{l}_{\nu, ab} s^{l}_{\nu} (r),
\end{align}
where $V^{l}_{\nu, ab}$ is a learned tensor parameter. The sensitivity functions $s$ are modified Gaussians, taking inverse pairwise distances as inputs, as in the expression below
\begin{align}
    s^{l}_{\nu} (r) = \mathrm{exp} \left [ - \frac{(r^{-1}-\mu^{-1}_{\nu,l})^2}{2\sigma^{-2}_{\nu,l}} \right] \varphi\sub{cut} (r),
\end{align}
where $\mu_{\nu,l}$ and $\sigma_{\nu,l}$ are learned parameters. The cutoff function $\varphi\sub{cut}$ is 
\begin{align}
    \varphi\sub{cut} (r) = 
    \begin{cases}
        \left[ \mathrm{cos} \left( \frac{\pi r}{2 R\sub{cut}}\right) \right]^2 & r \leq R\sub{cut} \\
        0 & r > R\sub{cut}.
    \end{cases}
\end{align}
The user-defined $R\sub{cut}$ is illustrated in the form of shaded circles on the nanoparticle in the Fig.~\ref{fig:HIPNNModel}~(b). 

The layers in HIP-NN are organized into hierarchical blocks, each beginning with an interaction layer, followed by a number of on-site layers. Each atom's charge at time $t$ is predicted hierarchically by adding
the predictions from the last layer $l_n$ of each interaction block $n$,
\begin{align}
    \delta \hat{q}_i(t) = \sum_n \delta \hat{q}^{(n)}_i (t) = \sum\super{N\sub{features}}\sub{a=1} \omega^{(n)}_{a} z^{l_{n}}_{i,a} + b^{(n)}.
\end{align}
This term is shown in red in Fig.~\ref{fig:HIPNNModel}~(b). Terms at higher hierarchical order $(n)$ are capable of taking into account effects from more distant neighbors.

\subsection{Single-step training}
\label{subsec:training1}

Training the network consists of adjusting the parameters ($W$, $V$, etc.) to improve the prediction of $\delta q_i(t)$. This is accomplished by minimizing a loss function quantifying network error. The total loss function is a combination of the mean absolute errors (MAE) and root mean square errors (RMSE) of $\delta q_i(t)$, as well as regularization terms which help to improve the generality of the final trained model. The total loss to minimize is
\begin{align}
\Lagr = \frac{1}{\sigma} (\mathrm{MAE} + \mathrm{RMSE}) + \Lagr\sub{L2} + \Lagr\sub{Q\sub{cons}},
\end{align}
where $\sigma$ is the standard deviation of $\delta q (t)$ in the training set. The term $\Lagr\sub{L2}$ regularizes the weights in both the on-site, interaction, and prediction layers and is defined as
\begin{align}
\Lagr\sub{L2} = \lambda_{L2} \left ( ||\omega||^{2}_2 + || W||^{2}_{2} + ||V||^{2}_{2} \right ),
\end{align}
where $\lambda_{L2}$ is the regularization weight. Additionally, we postulate that the nanoparticles are charge neutral, and enforce this charge conservation in loss as
\begin{align}
\Lagr\sub{Q\sub{cons}} = \lambda\sub{Q\sub{cons}} \left < \sum^{N\sub{atoms}}_{i} \delta \hat{q_i}(t) \right >^2_{D}.
\end{align}
The weight $\lambda\sub{Q\sub{cons}}$ controls the relative strength of the conservation term in the total loss.

These models contain a number of hyper-parameters, that is, model and training parameters which cannot be learned via gradient descent, but rather have to be explored using validation data. We perform this exploration as detailed in the Supporting Information (SI) in Ref.~\citenum{SIpaper}. Ultimately, the network architecture we used to train our models consists of three interaction layers, each followed by three atomic layers. The width of each layer is 30 neurons, and 10 sensitivity functions are
used with cutoff radius of minimum 1.5~$\text{\AA}$ and maximum 10~$\text{\AA}$. The cost function described above is minimized using ADAM variant of stochastic gradient descent.\cite{Kingma2014ADAM} The regularization parameter $\lambda_{L2}$ is set to 10$^{-6}$ and 10$^{-5}$ for Mulliken charges and electrons/holes, respectively. The charge conservation regularization parameter $\lambda\sub{Q\sub{cons}}$ is set to 10$^{2}$ for induced Mulliken charges and 1 for electrons/holes. We scale $\delta q (t)$ by a factor of $10^2$ and electron/holes by $10^4$ to ensure the network is predicting order one quantities. The model is trained to 60\% of the data. On every epoch in the training, the model is validated on 20\% of the data. To prevent overfitting, this process continues until the validation error stops improving for a number of epochs (the \emph{patience} for training), which is set to 50. After this, the learning rate is decreased by a factor of 2 and the training resumes with this new learning rate. Training terminates when $100$ epochs are completed with no improvement to the validation loss. For all the models presented here, the training ends after $\approx$10000 epochs and when the initial learning rate of 10$^{-3}$ has decreased to $\approx$10$^{-8}$. These parameters are determined by empirical testing detailed in the SI.\cite{SIpaper}

\subsection{Multi-step training}
\label{subsec:recurrTrain}

\begin{figure*}
\includegraphics[width=12cm, height=6cm]{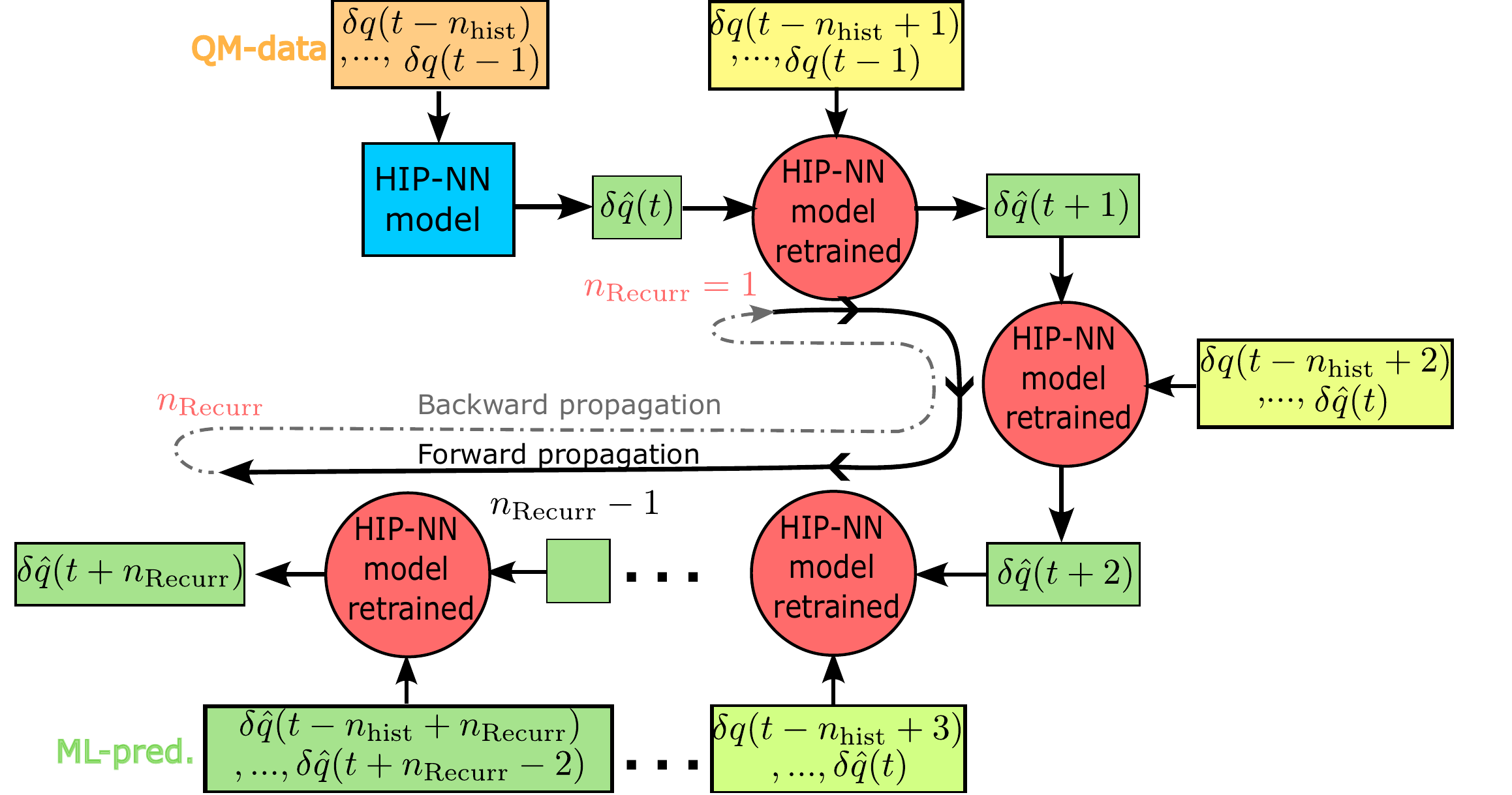}
\caption{Schematic of recurrent training model employed in this work. The trained HIP-NN model (blue square box) predicts charges at a single time point (green boxes). These predicted charges at time $t$ are added into the inputs in the first iteration of the recurrent training model (red circles) to predict charges at time $t+1$ and save associated losses. The method continues predicting future charges and saving their losses for free-parameter $n\sub{Recurr}$ where eventually all inputs into the model are ML predicted (visually indicated using the color gradient of the input boxes). The model parameters are then updated using the total loss accumulated over the predicted trajectory.}
\label{fig:recurrentSchem}
\end{figure*}

The prediction of a trajectory of charges over time entails feeding the prediction of the model at a current time $t$ as part of the input for the model at time $t+1$, where a full trajectory consists of 1300 time steps. Single-step prediction does not account for the possibility of compounding errors over time, as output charges at one step influence all subsequent steps. To address this, we explore multi-step training, which entails a loss function based on applying the model, including its own predictions as input, over many time steps. Figure~\ref{fig:recurrentSchem} contains the workflow diagram for this new method of training. The procedure starts with the prediction of $\delta q(t)$ given QM-calculated input charges $\left \{\delta q(t-n\sub{hist}), ... , \delta q(t-1)\right \}$ using a single-step trained model (blue box). In the next iteration, we shift the input charge history to remove the oldest point, and add in the new predictions, so that the model is fed with $\left \{\delta q(t-n\sub{hist}+1),..., \delta q(t-1), \delta \hat{q} (t)\right \}$ as input and predicts $\delta q(t+1)$. This process is repeated to generate a trajectory of predictions from the model for a number of steps $n\sub{Recurr}$ that is variable. When $n\sub{Recurr} > n\sub{hist}$, the inputs to the model are all ML-predicted charges from previous iterations. In Fig.~\ref{fig:recurrentSchem}, this is visually indicated by a color gradient for input boxes, from yellow to green. To accomplish multi-step training, we also compute losses and accumulate them into a total trajectory-based loss at each iteration, that is,
\begin{equation}
\Lagr\sub{multi-step} = \sum_{\tau=0}^{n\sub{Recurr}} \Lagr\sub{single-step}(t+\tau)    
\end{equation}
After a full $n\sub{Recurr}$ iterations, we call the ADAM optimizer to back-propagate errors from the final step all the way back to the first inference. This method of training, while significantly more computationally demanding per network update, may more closely capture the physics of charge propagation, because updates to the model take into account an error accumulated in the input charges over time.

\section{Results \& Discussion}

\subsection{Modeling Mulliken charge propagation}

\subsubsection{Single-step charge prediction}
We start by presenting the results of training the ML model to Mulliken charges per atom at individual time points. We have performed several tests varying HIP-NN hyper-parameters to find a robust model as detailed in the SI,\cite{SIpaper} resulting in two models, one with a short history parameter $n\sub{hist}=3$ (Model 1), and another with a longer history parameters, $n\sub{hist}=9$ (Model 2). The performance of these models is detailed in Table~\ref{tab:error}, showing the RMSE and MAE errors of predictions for time $t$ given the quantum-mechanical ground truth for the $n\sub{hist}$ time points prior. This is shown on a held-out test set of time frames from the training systems, Ag$\super{Icos}\sub{55}$, Ag$\super{Octa}\sub{146}$, and Ag$\super{Icos}\sub{147}$, as well as frames from Ag$\super{Icos}\sub{309}$ and Ag$\super{Icos}\sub{561}$ nanoparticles which constitute extensibility tests, that is, the systems are far larger than those used in training. These models are evidently quite robust, as the errors are far less than $10^{-5} e$ in all cases. On the larger nanoparticles Ag$\super{Icos}\sub{309}$ and Ag$\super{Icos}\sub{561}$, while the performance is still good, it is noticeably degraded in comparison to the smaller nanoparticles, particularly in terms of the RMSE, which can be about 6.8 times higher than the smaller nanoparticle test set. In contrast, MAE performance degrades by a factor of about 3.3 at most. The larger ratio between RMSE and MAE for the extensibility sets indicates that there are some outlier predictions in these sets. In other words, there are heavier tails in the extensibility set error distribution in comparison with the test set.

\begin{table}
\caption{Single-step trained HIP-NN model errors (RMSE and MAE) reported for held-out test sets for the nanoparticles included in the training, Ag$\super{Icos,Octa}\sub{55...147}$, and the larger ones not used for training, Ag$\super{Icos}\sub{309, 561}$. Values in the table are given in units of $10^{-5} e$.}
\label{tab:error}
\begin{center}
\begin{tabular}{ |p{1.4cm}|p{1.6cm}|p{1.4cm}|p{1.2cm}|p{1.cm}| } 
\hline
Model & Nano-particle & Trained on & RMSE & MAE \\
\hline
\hline
\multirow{3}{4em}{Model 1, $n\sub{hist}$=3} & Ag$\super{Icos,Octa}\sub{55...147}$ & Yes & 0.032 & 0.017 \\ 
\cline{2-5}
& Ag$\super{Icos}\sub{309}$ & No & 0.175 & 0.057 \\ 
\cline{2-5}
& Ag$\super{Icos}\sub{561}$ & No & 0.22 & 0.027 \\ 
\hline
\multirow{3}{4em}{Model 2, $n\sub{hist}$=9} & Ag$\super{Icos,Octa}\sub{55...147}$ & Yes & 0.042 & 0.022 \\ 
\cline{2-5}
& Ag$\super{Icos}\sub{309}$ & No & 0.19 & 0.066 \\ 
\cline{2-5}
& Ag$\super{Icos}\sub{561}$ & No & 0.062 & 0.023 \\
\hline
\end{tabular}
\end{center}
\end{table}

\subsubsection{Multi-step charge prediction}

The overarching aim of these models is to eventually replace expensive electron dynamics simulations with ML inferences, enabling rapid simulations of charge dynamics for tens of femtoseconds. While excellent single-step prediction shows an interesting capability for ML models to capture some aspects of charge relaxation physics, it does not address surrogate simulation capacity, because the input to these predictions is itself a series of expensive QM calculations. 
To test the ability to use these models as a surrogate for simulation, we apply the model in a multi-step fashion, folding the model predictions at time $t$ into the inputs for a prediction at time $t+1$ for an entire 1300 step trajectory. This probes the extent to which compounding errors affect the model performance in the prediction of plasmonic relaxation trajectories. Since the goal of our ML modeling is to create a faster prediction than expensive many electron QM, we first measured the approximate time to make predictions. Without particular attention to efficiency, for Ag$\super{Icos}\sub{55}$ ML model predicts the 1300 time steps trajectory in approximately 30 seconds on a 16 core CPU, taking less than 25~ms per time step. In contrast, this is 200 times faster than the cheapest rt-TDDFT simulation, which takes about 5 seconds per time step on 16 cores. More importantly, this speed-up improves by an order ($\sim$ 4000$\times$ as compared with rt-TDDFT) when the number of atoms are 10 times larger in the system (e.g., for Ag$\super{Icos}\sub{561}$). 

We compare the multi-step predictions from ML models to the ground truth QM simulations. In QM, given the plasmon resonance from the applied pulse, the charges oscillate over time until they decay through the Landau damping mechanism.\cite{Kawabata1966SingleParticlePlasmonCoupling, Shahbazyan2016LandauDampingInNano} This damping mechanism is not a smooth exponential function in nanoparticles of small sizes as a result of the discrete electronic energy spectrum. Instead, Rabi-oscillation-like characteristics show up in the dynamic dipole moments over a few tens of femtoseconds. These features are attributes of energy transfer between on-resonant single particle excitations creating electron-hole pairs and off-resonant excitations that constitute the plasmon modes.\cite{Ma2015AtomisticRttddftSilver55} For brevity, in the rest of this paper, we refer to these oscillations as recurring plasmons.  As a result of the oscillatory behavior of the system, we use the average autocorrelation $f(\Delta t)$ to summarize the dynamics of QM and ML charges across all atoms in the system:

\begin{multline}
    f(\Delta t) = \frac{1}{N\sub{atoms}}\sum_{i}^{N\sub{atoms}} \\
     \frac{\sum_{t=0}^{(t\sub{max}-\Delta t)} (\delta q^i_t - \bar{\delta q}^i) (\delta q^{i}_{t+\Delta t} - \bar{\delta q}^i)}
    {\sum_{t=0}^{t\sub{max}} (\delta q^{i}_t - \bar{\delta q}^i)^2}.
\end{multline}
This measure not only includes all atoms, but also normalizes each atom's contribution by its variance; the importance of each atom in the average autocorrelation is independent of the magnitude of charges on that atom.

\begin{figure}
\includegraphics[width=\columnwidth]{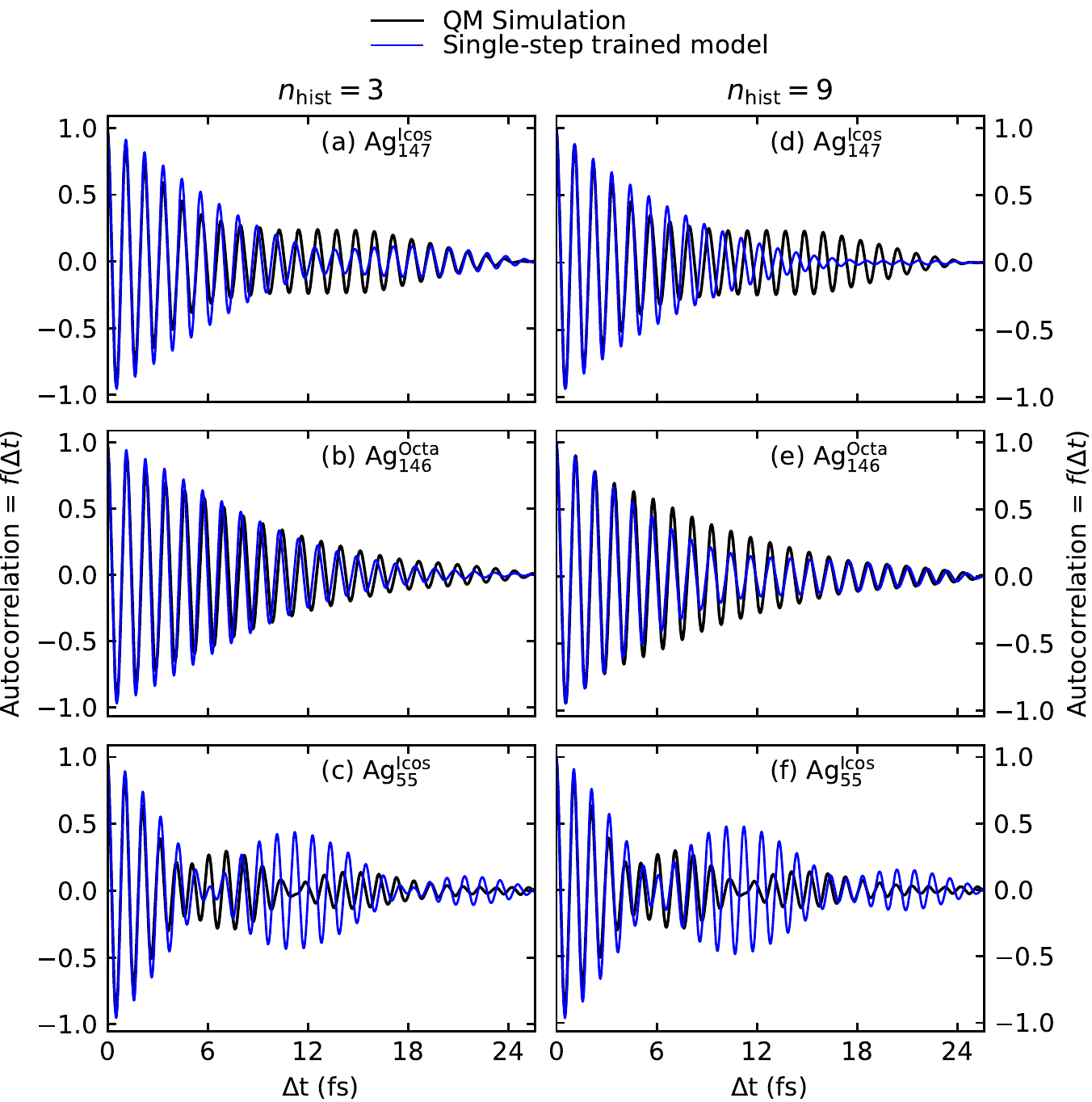}
\caption{Comparison of predicted vs. QM-simulated autocorrelation of the charges over time for ML model 1 ($n\sub{hist}$ = 3) and model 2 ($n\sub{hist}$ = 9). The ML-predicted results belong to a test trajectory in held-out test sets for each of the nanoparticles.}
\label{fig:ACNhist}
\end{figure}

The comparison of the charge dynamics as predicted by HIP-NN with the ground truth GPAW simulations are presented as autocorrelation plots in Fig.~\ref{fig:ACNhist}(a-f), which compares the ML-predicted (blue) autocorrelation with the QM-calculated (black) autocorrelation for the two  models with $n\sub{hist}$=3, 9 and for  three types of nanoparticles,  Ag$\super{Icos}\sub{55}$, Ag$\super{Oct}\sub{146}$, Ag$\super{Icos}\sub{147}$. For each nanoparticle type, we show one randomly selected trajectory from the test set.

 In these plots, the first striking feature is that ML models predict autocorrelations that match well with QM simulations up to time lags of approximately 6~fs for all nanoparticles. Keeping in mind that the dynamic charge data is generated with a time step of 20 attoseconds, this represents 300 time steps or sequential ML inferences. This indicates robust numeric stability of the HIP-NN model. It also demonstrates the possibility of predicting charge dynamics in these systems with only $n\sub{hist}$ = 3. In fact, the models for both $n\sub{hist}$ = 3 and $n\sub{hist}$ = 9 show similar results, indicating that there is no strong dependency on the amount of charge history used in making a prediction. That is desirable feature to accelerate dynamics calculations with minimal expensive QM calculation input. Note that the agreement is excellent for the octahedron Ag$\sub{146}$ for the entire simulation time. This could hint that the plasmons interact less with electron-hole pairs for a few 10s of femtoseconds in octahedron nanoparticles. For icosahedral nanoparticles, the agreement is better for the larger icosahedral nanoparticle Ag$\sub{147}$ than for the smaller Ag$\sub{55}$. For Ag$\sub{55}$, the ML predictions capture the dynamics at short timescales $<6$~fs, but mispredict the multiple recurrence phenomena in reference data, which do appear in ML, but at larger time-scales and with larger amplitudes.

\begin{figure}
\includegraphics[width=\columnwidth]{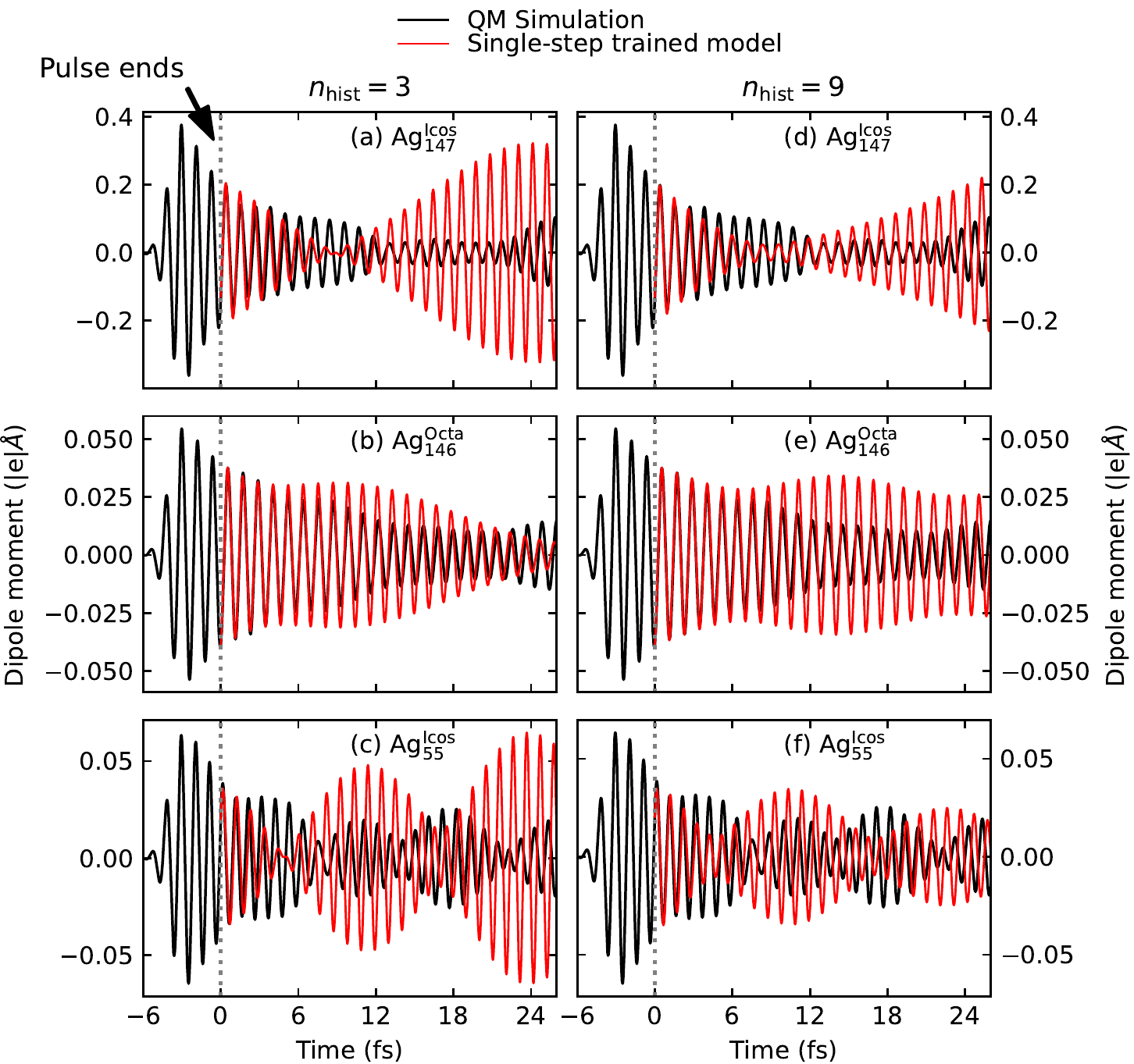}
\caption{Comparison of predicted vs. QM-simulated dipole moments in $x$ direction over time for the two selected ML models with $n\sub{hist}$ = 3, 9. The ML-predicted dipoles starting at $t$ = 0~fs belong to a test trajectory in the held-out test sets for each of the nanoparticles.}
\label{fig:dipoleNhist}
\end{figure}

Figure~\ref{fig:dipoleNhist}(a-f) presents a comparison of ML-inferred (red) and QM-simulated (black) dipole moments for the test trajectories previously examined for each nanoparticle type in the training set.  Remarkably, we see that with only $n\sub{hist}$ = 3, 9 data points from QM data, both models are able to predict dipole moments in strong quantitative agreement with QM results up to $\sim$5~fs (250 time steps) after the pulse ends.  As discussed briefly before, the shape of a nanoparticle affects Landau damping and introduces plasmon recurrence. The octahedron Ag$\super{Oct}\sub{146}$ does not show any strong recurrence. On the other hand, both icosahedral shaped Ag$\super{Icos}\sub{55}$ and Ag$\super{Icos}\sub{147}$ have complex dipole moment profiles with multiple plasmon recurrences. Interestingly, this recurrence is stronger when the nanoparticle is small (in Ag$\super{Icos}\sub{55}$) as compared with a large one (Ag$\super{Icos}\sub{561}$ (see Fig.~S9). These recurrences likely originate from the complex convolutions of wave function phases leading to long-time coherent dynamics. Letting the simulation run for a long period ($>$ 50~fs) might exhibit a steady decay of the dipole moment (as shown in Fig.~S9 in SI.\cite{SIpaper} Critically, the ML models capture these qualitatively different behaviors in the plasmon resonances, even without being trained on the phase information in Kohn-Sham wave functions. This agreement is most evident in  Ag$\super{Icos}\sub{55}$, when the HIP-NN models capture multiple plasmon oscillations. 
\begin{figure}
\includegraphics[width=\columnwidth]{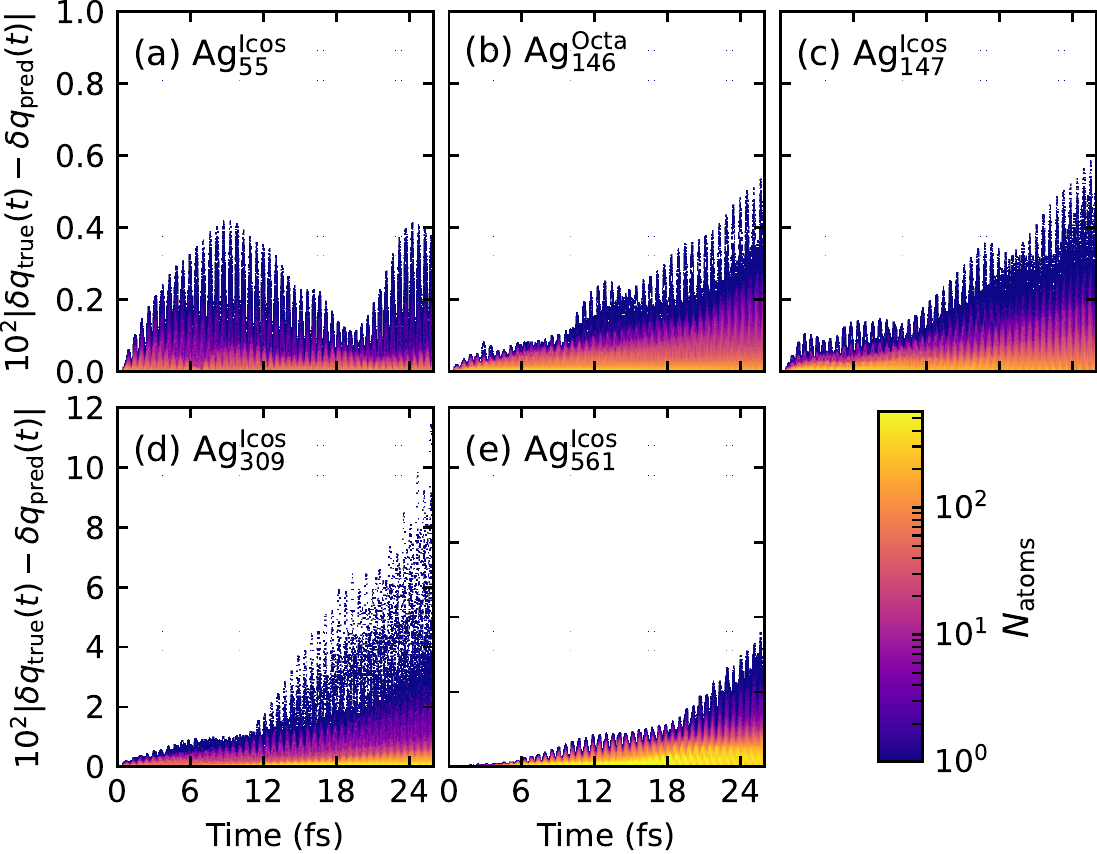}
\caption{Computed absolute error of ML-inferred charges vs. true charges over time and histogrammed by number of atoms (as shown in the colorbar) using $n\sub{hist}$ = 9 model. (a-c) shows errors calculated for nanoparticles that were used in the training. (d,e) presents errors between ML-inferred and true charges in nanoparticles. For each nanoparticle, the held out test set contains (20-25) trajectories.}
\label{fig:chargeInferErr}
\end{figure}
To get a broader statistical picture of multi-step ML inference performance plasmonic dynamics, Fig.~\ref{fig:chargeInferErr} (a-e) shows absolute error between true and ML-inferred charges for held-out test sets of each nanoparticle type, with 20-25 trajectories per nanoparticle type. The errors are shown as a series of histograms as a function of time; no aggregation is performed in the time dimension. As an extensibility test, we also apply the HIP-NN models to Ag$\super{Icos}\sub{309}$ and Ag$\super{Icos}\sub{561}$, which are much larger than the nanoparticles included in the training data set. A common feature in most of the nanoparticles is that a few atoms ($< 10$) show large errors, particularly after $\sim$5~fs, except for Ag$\super{Icos}\sub{55}$ which exhibits a tight error distribution even at 25~fs. This could be related to the stronger plasmon recurrence in this nanoparticle. 
Interestingly, for all nanoparticles with geometry represented in the training set, more atoms have very low errors ($<10^{-4}e$) until $\sim$25~fs (Fig.~\ref{fig:chargeInferErr} (a-c)). On the
other hand, for the extended nanoparticles of Ag$\super{Icos}\sub{309}$ and Ag$\super{Icos}\sub{561}$, the errors are larger by an order of magnitude. For all cases, including extensibility nanoparticles, ML predictions perform very accurately for the bulk of the atoms, but with a tail in the absolute error distribution, showing that a few atoms are mispredicted. This tail grows steadily over time in all geometries besides Ag$\super{Icos}\sub{55}$.

\subsubsection{The effects of multi-step training}
\begin{figure}
\includegraphics[width=\columnwidth]{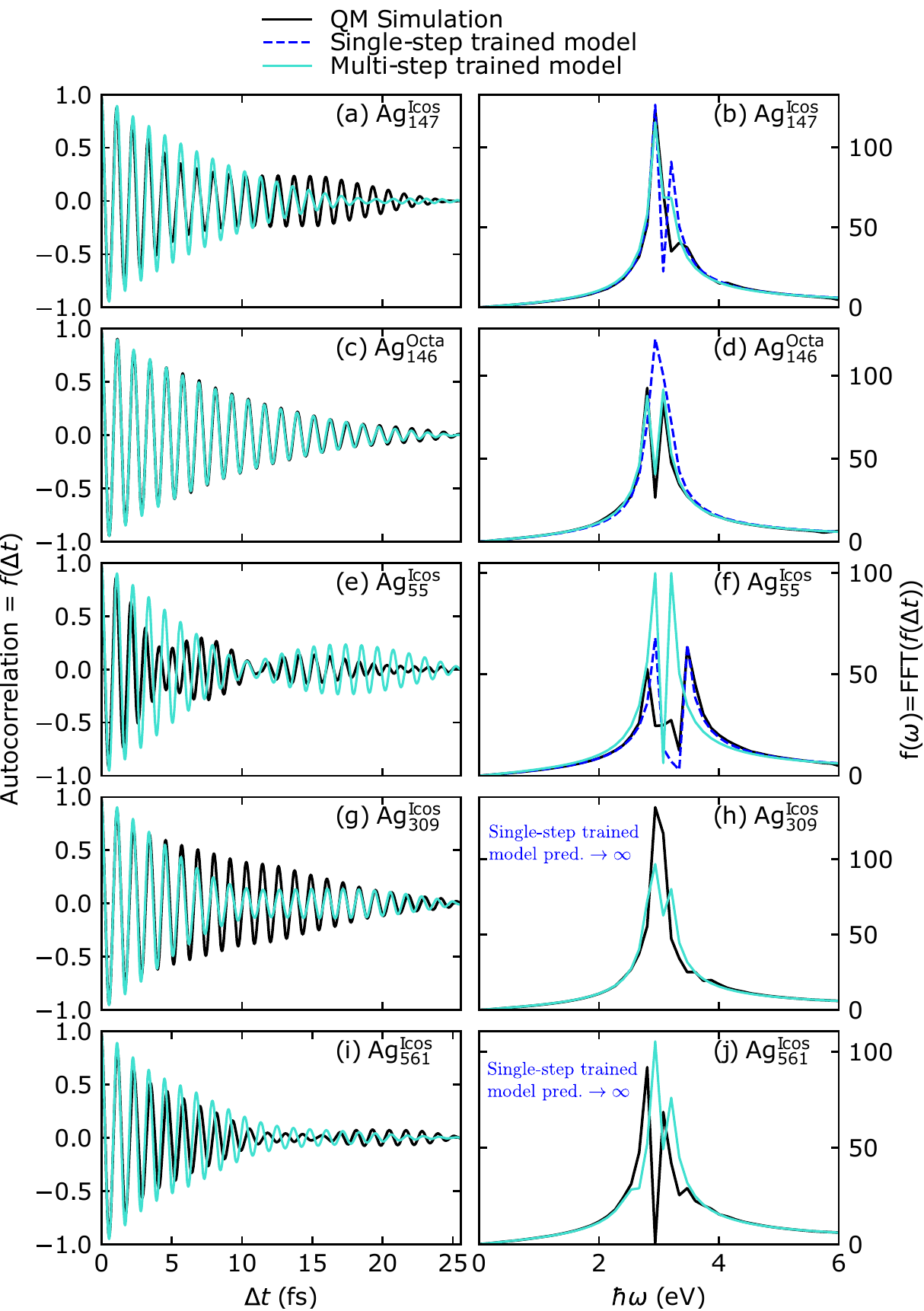}
\caption{Comparison of the single-step-trained HIP-NN (blue) and the multi-step-trained HIP-NN (turquoise) induced Mulliken charge autocorrelations for a sample trajectory for all nanoparticles, including Ag$\super{Icos}_{55-561}$ and Ag$\super{Octa}\sub{146}$. The model takes $n\sub{hist}$ = 3 and it is trained in multiple steps with $n\sub{Recurr}$ = 800 future time steps in the trajectory. Column 1 in the panel shows the Mulliken charge autocorrelation functions for all the nanoparticles against the QM-calculated, and column 2 shows their Fourier transforms as functions of frequency (energy). For the extended nanoparticles of Ag$\super{Icos}\sub{309}$ and Ag$\super{Icos}\sub{561}$, single-step HIP-NN model predictions become numerically large after only $\sim$200 time points as shown in Fig.~S6 in SI. So, it is not plotted in here, and only indicated by ``single-step trained predictions $\rightarrow \infty$" in panels (h,j).}
\label{fig:recurrAg55-561Charge}
\end{figure}

As discussed in the section~\ref{subsec:recurrTrain}, refine the ML models using multi-step training over $n\sub{Recurr}$ time steps to improve predictions over time. Figure~\ref{fig:recurrAg55-561Charge} (a-f) presents autocorrelation of the induced Mulliken charges and the Fourier transform thereof, calculated using $n\sub{hist}$ = 3 HIP-NN model trained multi-step wise for $n\sub{Recurr}$ = 800, for a sample trajectory of each nanoparticle in the training set. Since multi-step training starts from an already trained HIP-NN model, the change in the model is expected to be small. This appears to be true on the whole; for example, compare the autocorrelations in Fig.~\ref{fig:ACNhist} (a-c) with the HIP-NN-multi-step predicted ones in Fig.~\ref{fig:recurrAg55-561Charge}~(a,c,e). This can be further seen in Fig.~\ref{fig:recurrAg55-561Charge} (b,d,f), which shows the frequency content of the system explicitly. 
For Ag$\super{Icos}\sub{147}$, the autocorrelation has a diminished two-frequency signal correctly predicted by multi-step trained models, whereas single-step models exhibit a strong two-frequency structure. For Ag$\super{Octa}\sub{146}$, while this single-step-trained model predicts a frequency close to the two ground-truth frequencies, multi-step-trained model predicts two frequencies, and they are in excellent agreement with the QM-simulated frequencies. For Ag$\super{Icos}\sub{55}$, multi-step trained model appear to perform somewhat weaker than single-step trained model, underestimating the gap between the two frequency components.
Extended nanoparticles show an even more important improvement brought about during multi-step training.  Ag$\super{Icos}\sub{309}$ and Ag$\super{Icos}\sub{561}$ nanoparticles cannot be modeled over time using the single-step models, as the the predictions diverge after a a few fs of simulation (see SI\cite{SIpaper}). Figure~\ref{fig:recurrAg55-561Charge} (g-j) show autocorrelations and their Fourier transforms of induced Mulliken charges using the  multi-step $n\sub{hist}$ = 3  HIP-NN model. Remarkably, multi-step-training model stabilizes the model, producing predictions that are in excellent agreement with the ground-truth for $\sim$5~fs after the pulse ends (see Fig.~\ref{fig:recurrAg55-561Charge} (g,i)). For the rest of the trajectory, the predictions qualitatively match with the QM-data. This is visible in the frequency content, Fig.~\ref{fig:recurrAg55-561Charge} (h,j). 

\subsection{Modeling of hot carrier propagation}

So far, we have shown that ML models can quickly predict dynamical properties such as induced Mulliken charges and dipole moments over a few tens of femtoseconds. Plasmonic hot carriers may also be predicted, which would enable understanding the role of geometric factors that impact device efficiency.  In this case, the ML model then predicts two quantities per atom, and takes as input both the electron and hole history for the atoms.
\begin{figure}
\includegraphics[width=\columnwidth]{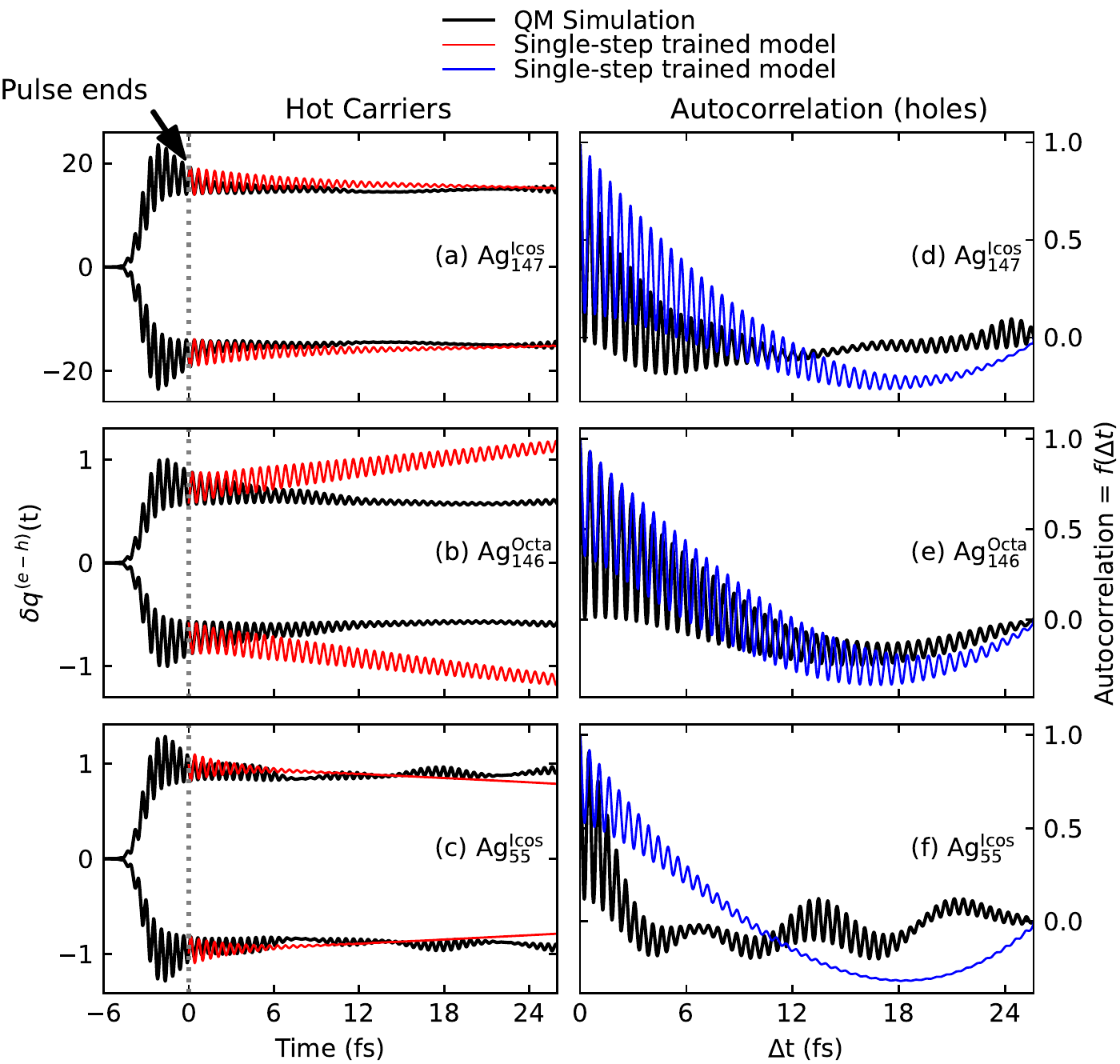}
\caption{Comparison of ML-predicted (blue and red) vs. QM-simulated (black) positive and negative charges and autocorrelation over time for an ML model with $n\sub{hist}$= 9. (a-c) shows sum of the positive charges (hot holes) and negative charges (hot electrons) over all atoms in the nanoparticle over time. (d-f) shows autocorrelation for hot holes only (hot electrons have a similar trend). The ML-predicted results belong to a test trajectory in the held-out test sets for each of the nanoparticles.}
\label{fig:heACNhist}
\end{figure}
Figure~\ref{fig:heACNhist} (a-c) presents ML-predicted (red) against QM-simulated (black) population probabilities of both electrons and holes. Figure~\ref{fig:heACNhist} (d-f) shows the autocorrelation functions for the holes only; electrons have a similar trend and are therefore omitted. First, the underlying QM-calculation shows that the population probabilities of hot carriers appears to settle towards constant value over time up to $\sim$32~fs. This is expected because the rt-TDDFT dynamics simulation only accounts for the optical-single-particle interaction, and any scattering event such as e-e or e-ph that can equilibrate carriers is missing. Nonetheless, what we can infer from these probabilities is that in larger nanoparticles, Ag$\super{Icos}\sub{147}$ and Ag$\super{Octa}\sub{146}$, the amplitudes of the oscillations are decaying more steadily as compared to the smaller Ag$\super{Icos}\sub{55}$. For Ag$\super{Icos}\sub{55}$, the amplitude shows a more complex profile with recurring features, similar to what is observed in the dynamic dipole profile. The ML model for these quantities is qualitatively accurate for all nanoparticles; electron and hole densities remain near a baseline, around which there is an oscillation with varying amplitude in time. This is visible in autocorrelation plots of Ag$\super{Icos}\sub{147}$ and Ag$\super{Octa}\sub{146}$ in Fig.~\ref{fig:heACNhist} (d,e), where the ML-predicted lines (blue) follow the same frequency pattern as the underlying QM-simulated (black) for almost 32~fs. 
For the Ag$\super{Icos}\sub{55}$, the agreement declines after a few femtoseconds in the beginning (Fig.~\ref{fig:heACNhist} (f)) but it qualitatively predicts the convergence of both hot electrons and holes towards a constant value at time $\sim$25~fs (Fig.~\ref{fig:heACNhist} (a,c)).  Thus, ML prediction of hot carrier dynamics appears to be noticeably more difficult than modeling charge dynamics.

\section{Conclusion}

In this work, we have extended an atomistic neural network, HIP-NN, to simulate plasmon dynamics. We do this by feeding the charge history before and including time $t$ into the network and predicting the charge at time $t+1$. Networks are trained and tested on silver nanoparticles with icosahedral and octahedron shapes of up to 147 atoms, as well as tested on larger nanoparticles, up to 561 atoms, to assess the robustness of the model. The resulting model captures important phenomena, including plasmonic resonance and decay into hot carriers, is stable for the entire 1300 time-step (32 fs) trajectory, and is at least 200$\times$ faster than the reference rt-TDDFT method. To address compounding errors over time, we also explored multi-step training to charge trajectories, which allows refining the model through a loss function, which accounts for many steps of prediction in a row. 

In particular, for Mulliken charges, we show that the models are extremely quantitatively accurate when making single-step predictions, with typical errors less than $10^{-5} e$. In trajectory prediction, charges are remarkably accurate for approximately 5~fs. After this time, compounding errors play a stronger role, but model accuracy is generally still better than $10^{-3} e$. On larger nanoparticles, the model is not stable for trajectory predictions. However, we find that multi-step training stabilizes these predictions, and generally improves the performance on the smaller nanoparticles, as well, by examining the dynamic dipole moment and the autocorrelation of charges. The accuracy of these models is remarkable in light of the information lost by representing the system state using only Mulliken charges, which ignore both relative phase factors and atomic-orbital-level population information in the wave function. While in principle this missing information could have proved an obstacle to building a machine learning model, this work shows that short-time charge histories provide a large amount of information about the non-equilibrium state of the system, from which a surrogate model can be built.

Besides Mulliken charges, we also examine modeling of hot carrier dynamics using an electron/hole decomposition. The simulations are accurate for a few femtoseconds of prediction, but afterwards retain only qualitative accuracy.
The poorer performance when propagating hot carriers shows that there is still significant work to be done. It is very possible that ascribing electrons and holes to atoms is too coarse of a resolution for the ML model to capture the physical processes governing hot carrier dynamics.

We anticipate that further improvements in the quantitative prediction of long-time system behavior (such as plasmonic recurrence time and amplitude or hot carrier dynamics) may require adding back in information from the wave function (such as phase factors) that is neglected when using the Mulliken charge or electron/hole decomposition. For example, we hypothesize that better results might be obtained if the ML model processes hot carrier density in a much larger basis of atomic orbitals. However, atomic orbitals are not rotationally invariant. As such, an equivariant machine learning model will be required.\cite{Batzner2022Nequip, Schutt2019SchNOrb} These models can enable full density matrix predictions, and allow for the ML estimation of hot carrier probability distributions as a function of energy and time. This would be another important step towards realizing the potential for ML to predict long-time-scale hot carrier dynamics with enough throughput to fully explore the massive design space of plasmonic hot carrier devices.

\section*{Acknowledgments}
The work at Los Alamos National Laboratory (LANL) was supported by the LANL Directed Research and Development Funds (LDRD) and performed in part at the Center for Nonlinear Studies (CNLS) and the Center for Integrated
Nanotechnologies (CINT), a US Department of Energy (DOE)
Office of Science user facility at LANL. A.H. and N.L. acknowledge financial support from the LANL Laboratory Directed Research and Development program. B.N. and S.T. acknowledge support from the US DOE, Office of Science, Basic Energy Sciences, Chemical Sciences, Geosciences, and
Biosciences Division under Triad National Security (Triad) contract grant number 89233218CNA000001 (FWP:LANLE3F2). This research used resources provided by the
LANL Institutional Computing Program. LANL is managed by
Triad National Security for the US DOE’s NNSA, under contract number 89233218CNA000001. 

\bibliography{references}

\end{document}